\documentclass[conference]{IEEEtran}
\IEEEoverridecommandlockouts
\usepackage{amsmath,amssymb}
\usepackage{booktabs}
\usepackage{graphicx}
\usepackage[hidelinks]{hyperref}
\usepackage{url}
\usepackage{balance}

\title{When Routes Run Out: Adversarial Co-Learning and Explainable Robustness in Quantum Repeater Networks}
\author{
\IEEEauthorblockN{Brennan Bell\IEEEauthorrefmark{1},
Inti Gabriel Mendoza Estrada\IEEEauthorrefmark{2},
Andreas Tr\"ugler\IEEEauthorrefmark{3}, and
Paul Erker\IEEEauthorrefmark{4}}
\thanks{\IEEEauthorrefmark{1}RFI-IRFOS and TU Graz, Graz, Austria; bell.brennan.p@gmail.com}%
\thanks{\IEEEauthorrefmark{2}openmaind FlexCo and TU Graz, Graz, Austria; inti.mendoza@openmaind.ai}%
\thanks{\IEEEauthorrefmark{3}Know Center Research GmbH and University of Graz, Graz, Austria; andreas.truegler@uni-graz.at}%
\thanks{\IEEEauthorrefmark{4}Atominstitut, TU Wien and IQOQI, \"OAW, Vienna, Austria; paul.erker@tuwien.ac.at}%
}

\begin{document}
\bstctlcite{IEEEexample:BSTcontrol}
\maketitle

\begin{abstract}
We study an adversarial bandit problem for entanglement-based quantum-network routing over a modest graph corpus. Alice selects an end-to-end repeater route for an Ekert-91 protocol (E91) representing her move, while Eve selects an attack surface, either edge intercept--resend or repeater memory degradation. Payoffs are drawn from cached SeQUeNCe-simulated  E91 transcripts, and Alice accepts a turn when the finite-sample statistic violates the Clauser-Horne-Shimony-Holt (CHSH) bound. Performing adversarial co-learning across 50 structured topologies, we find that learned retention tracks a full-matrix minimax reference closely (Pearson $r=0.99$): under a one-surface Eve action model, bottleneck families have zero retention, while non-bottleneck families follow a $1-1/N$ coverage principle. We then fit decision-tree explanation models to graph-, attack-, and route-level topology-corpus targets and report their faithfulness. Finally, we construct prompt records for local language models to summarize the tree evidence, resulting in an open-source explanation workflow for quantum-repeater network games.
\end{abstract}


\section{Introduction}
\subsection{Research questions}
The present works analyses two main questions. Can a standard adversarial bandit, playing from bandit feedback with no topology knowledge, recover the strategic structure of route selection and attack placement in a simulated Ekert-91 (E91) repeater network? Can the learned and reference strategies then be \emph{explained}---by symbolic representations and small local language models---with measured faithfulness and without security-oriented hallucinations?

\subsection{Motivation}
In E91, Alice and Bob consume entangled pairs and use Bell-test statistics as a security monitor \cite{Bell66,Ekert1991, CHSH1969}. On a repeater network, route choice also induces a classical interdiction game: Alice's mixed strategy determines which fibers and memories are exposed, while a limited Eve chooses one component to attack. This route-versus-surface structure is well known in zero-sum network interdiction \cite{WashburnWood1995}. In the homogeneous case, a component on every Alice--Bob route exposes a fatal bottleneck to Eve, i.e. with $N$ component-disjoint routes and one attacked surface, uniform routing gives hit probability $1/N$, hence a retention of $1-1/N$.

The  question is therefore not whether an algorithm can rediscover this coverage logic in a binary graph game. It is whether the same strategic structure remains visible when payoffs are generated by topology-corpus E91 transcripts: CHSH-gated acceptance, shared repeater memories, unattacked failures, and component attacks. Explanations matter because simulator diagnostics are easy to over-interpret as quantum key distribution (QKD) security claims. 

\subsection{Related work}
SeQUeNCe and NetSquid make quantum-network hardware and control assumptions explicit \cite{Wu2021Sequence, Coopmans2021NetSquid}; recent trapped-ion network nodes motivate high-efficiency parameter scales \cite{Cui2026}. \emph{Exponential-weight algorithm for Exploration and Exploitation} (Exp3) is the standard adversarial bandit \cite{Auer2002Exp3}; in zero-sum games, no-regret learning supports time-averaged strategies, while terminal multiplicative-weight iterates may fail to converge \cite{FreundSchapire1999, BaileyPiliouras2018MWU}. For explanation, we use shallow decision trees with reported target fit \cite{Breiman1984CART, Rudin2019Interpretable}.

\begin{figure}[htb]
\centering
\includegraphics[width=\linewidth]{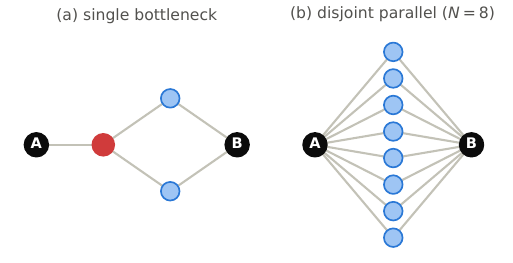}
\caption{Exemplar topologies. (a) In a single-bottleneck graph every route crosses one node (red): Eve can deny all traffic from one surface. (b) In a disjoint-parallel graph one attack surface covers only one of $N$ routes.}
\label{fig:exemplars}
\end{figure}

\section{Modeling and Methodology}

\subsection{Quantum repeater network simulation}
SeQUeNCe \cite{Wu2021Sequence} is a discrete-event simulator of quantum networks with explicit sources, fibers, memories, and control protocols; SeQUeNCe's repeater module is Barrett--Kok entanglement generation \cite{BarrettKok2005} plus entanglement swapping. In our E91 game turn, a budget of 350 entangled pairs is distributed end-to-end over the chosen route; delivered pairs are measured at random E91 settings, a subset is used for the Bell statistics to compute $S$ \cite{CHSH1969}, and matched basis pairs yield sifted key bits. The public outcome of a turn is one of \{accepted, CHSH abort, quantum bit error rate (QBER) abort, delivery failure\}, with the CHSH check preceding the QBER check.

\subsection{Minimax reference and Exp3}
For each graph, cached trials define Alice's payoff matrix $M\in[0,1]^{|\Pi|\times|\mathcal{E}|}$. Let $X=\Delta(\Pi)$ and $Y=\Delta(\mathcal{E})$. The full-matrix reference is
\begin{equation}
v^\star=\max_{x\in X}\min_{y\in Y}x^TMy,
\qquad
R_{\rm oracle}=\mathrm{clip}_{[0,1]}(v^\star/v_0),
\end{equation}
where $v_0$ is the best no-attack value. We measure learned averaged strategies relative to the Nash equilibrium gap
\begin{equation}
g(\bar{x},\bar{y})=
\max_{x\in X}x^TM\bar{y}
-
\min_{y\in Y}\bar{x}^TMy ,
\end{equation}
which is zero at a saddle point of this empirical zero-sum game.

Both players use Exp3 \cite{Auer2002Exp3}: action probabilities mix normalized weights with uniform exploration, played rewards are importance-weighted, and weights are updated multiplicatively. We use a learning rate of $\eta_t=\sqrt{\log K/(K(t+10^4))}$ and an exploration parameter $\gamma_t=\min\{0.2,K\eta_t\}$, where $K$ is the action-space size and $t$ is the game-turn. Since no-regret guarantees concern averaged play, all reported strategies are frequency-of-play averages rather than terminal iterates \cite{FreundSchapire1999, BaileyPiliouras2018MWU}.

\section{Experiments}

\subsection{Implementation choices}
The corpus holds 50 graphs from eight structural families (single/multi/deep bottleneck, deep/disjoint/layered parallel, length-variant disjoint, Wheatstone chains), with up to 32 routes of at most 7 hops and edge lengths 400--550\,m. Fig.~\ref{fig:exemplars} shows the primary structural motifs. Eve's action set is $\{\text{no attack}\}\cup\mathcal{E}_{\rm ir}\cup\mathcal{E}_{\rm mem}$: one intercept--resend action per eligible edge and one degradation action per internal repeater memory. Intercept--resend is the eavesdropper whose intermediate measurements destroy the CHSH violation \cite{Ekert1991}, while memory degradation models repeater node integrity/availability disturbances \cite{Satoh2021}. Practical QKD attacks such as detector timing \cite{Qi2007TimeShift} and Trojan-horse probing \cite{Gisin2006Trojan} motivate component-specific threat modelling, but those detector/source side channels are not implemented in this experiment.

\begin{figure}[b]
\centering
\includegraphics[width=\linewidth]{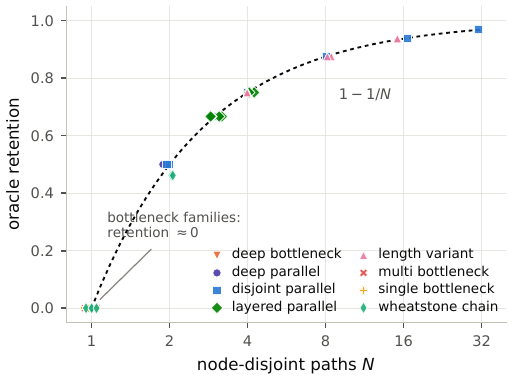}
\caption{Oracle retention versus node-disjoint path count $N$ for all 50 topologies ($\log_2$ axis; coincident points jittered). Non-bottleneck families track $1-1/N$; bottleneck families sit at zero.}
\label{fig:retention_paths}
\end{figure}

To make $5\times10^5$-turn co-learning feasible, Exp3 turns are sampled from caches of pre-simulated SeQUeNCe E91 trials. We store 64 trials per route--hop and per attack--hop profile; during training, we sample the matching pool, while the complete-information matrix uses the mean Alice acceptance over 16 cached draws per cell. Alice's payoff is \emph{CHSH-only}: after forming the Bell-test sample, a turn is accepted if and only if $|S|>2$. QBER is logged but not used in the payoff, since the CHSH test dominates the statistics in this simulation. Eve payoff is binary, whether or not her attack lies on Alice's route. 

We use a memory fidelity of $0.98$ and, based on recent multiplexed trapped-ion network results \cite{Cui2026}, a memory efficiency of $0.544$; however, we apply them to a SeQUeNCe Barrett--Kok model; this is a deliberate compromise between accurate device modelling and realistic repeater performance. A pre-run health check expects at least 319 of 350 entangled pairs to be delivered, where 4 of 9 combinations feed the CHSH test for Alice and 2 of 9 combinations feed the key bits. Without unattacked failures, the induced game would be zero-sum; CHSH failures occur on roughly a fifth of clean turns, making the game only approximately zero-sum.

\subsection{Control and target outcomes}
As a control, we aggregate the caches by route depth. Clean baselines violate the CHSH bound at every depth: mean $|S|$ falls from $2.75$ at one hop to $2.37$ at seven hops. Active hits, by contrast, never violate the CHSH bound. QBER adds no independent signal: any attack disruptive enough to corrupt the key already breaks the CHSH violation, and the Bell check occurs first.

The target outcome is the strategic structure. Fig.~\ref{fig:retention_paths} shows that the minimax reference is nearly determined by two features. Non-bottleneck families track the coverage principle $1-1/N$ in the node-disjoint path count $N$, while every bottleneck family collapses to zero retention: Eve parks on the unavoidable cut. Exp3 recovers this reference value. Final learned retention, measured as mean acceptance over the last $2.5\times10^4$ turns and normalized by $v_0$, matches oracle retention at Pearson $r=0.99$ across the 50 graphs. Five high-redundancy graphs sit slightly above one, with maximum ratio $1.11$, as an artifact of the unclipped finite-sample ratio.

\begin{figure}[t]
\centering
\includegraphics[width=\linewidth]{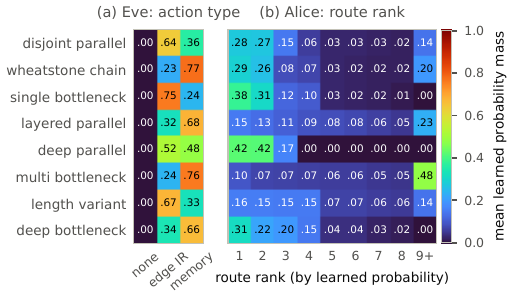}
\caption{Learned (time-averaged) Exp3 strategies aggregated by family. (a) Eve's probability mass by action type. (b) Alice's mass by route rank (routes sorted by learned probability per graph; ``9+'' sums the tail). Cell labels give the mean mass.}
\label{fig:heatmaps}
\end{figure}

Strategies in Fig.~\ref{fig:heatmaps} show Eve rarely plays no-attack, and her surface choice is structural: intercept--resend on the cut edge in single-bottleneck graphs, as in Fig.~\ref{fig:exemplars}a, and memory degradation in multi-bottleneck, deep-bottleneck, and Wheatstone families, where one shared repeater memory covers several routes. Alice mirrors this structure: she spreads nearly uniformly over disjoint routes, but learns a long-tailed hedge on the large multi-bottleneck route sets, with $0.48$ mass beyond rank~8.

\subsection{Connections to theory}
\begin{figure}[t]
\centering
\includegraphics[width=\linewidth]{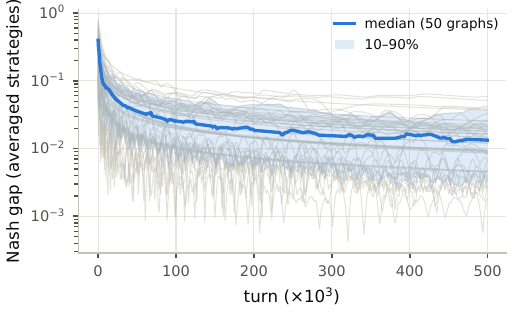}
\caption{Nash-gap decay across all 50 topologies (gray: individual graphs; blue: median and 10--90\% band). The median halves between $10^5$ and $5\times10^5$ turns.}
\label{fig:exploitability}
\end{figure}

The Nash gap of the time-averaged Exp3 strategies in Fig.~\ref{fig:exploitability} shows the median falls from $2.6\times10^{-2}$ at $10^5$ turns to $1.3\times10^{-2}$ at $5\times10^5$ turns, consistent with the no-regret view that zero-sum multiplicative-weight dynamics should be evaluated through averaged play rather than terminal iterates \cite{Auer2002Exp3, FreundSchapire1999, BaileyPiliouras2018MWU}. The non-monotonicity is empirical: payoff cells are estimated from discrete cached trials, best responses can switch under small changes in the average mixture, and some unattacked routes fail the finite CHSH test. The final gap grows with joint action-space size $K_AK_E$ (log--log $r=0.75$), but this is not topological vulnerability; high-redundancy non-bottleneck graphs retain the expected $1-1/N$ minimax value, while the residual gap measures horizon imbalance in the learned mixture.


\subsection{Decision-tree explanation models and LLM interpretation}
We fit depth-3 decision-tree regressors to topology-corpus targets; Table~\ref{tab:dt} reports their faithfulness, which spans the full quality range. The graph-retention tree is nearly exact ($R^2=0.9984$): its active rules use bottleneck presence and disjoint-path count, so graph-level tree summaries are faithful to our topology-corpus target. The Eve-action tree is moderately faithful ($R^2=0.7048$ over 1501 rows), led by target-route coverage and the memory-degradation flag, consistent with the coverage logic of Fig.~\ref{fig:heatmaps}a. The expected-denial and Alice-route trees are weak ($R^2=0.43$ and $0.24$), so route-level tree explanations capture the non-uniqueness of the oracle route target. Learned-target counterparts fitted for the prompt pack behave analogously (learned retention $R^2=0.98$, learned Eve strategy $R^2=0.68$).

These trees are the symbolic object digested by local large language models (LLMs) as part of an otherwise-static prompt requesting a response of at most $300$ words and structured into 5 sections: graph interpretation, tree evidence, strategy evidence, caveats, and rubric self-check. The local models were served via Ollama \cite{OllamaSoftware2026}: \texttt{llama3.1:8b}~\cite{OllamaLlama31}, \texttt{phi4:14b}~\cite{OllamaPhi4}, and ~\texttt{nemotron-3-super:120b}
\cite{OllamaNemotron3Super}.


\begin{table}[t]
\centering
\scriptsize
\setlength{\tabcolsep}{3.0pt}
\caption{Decision-tree faithfulness to the 50 corpus topologies.}
\label{tab:dt}
\begin{tabular}{@{}lcccc@{}}
\toprule
Target & Rows & Train $R^2$ & Held-out $R^2$ & Held-out MAE \\
\midrule
Oracle graph retention & 50 & 0.9984 & 0.9978 & 0.0081 \\
Oracle Eve action probability & 1501 & 0.7048 & 0.6886 & 0.0257\\
Expected denial vs. oracle Alice & 1501 & 0.4344 & 0.3781 & 0.1305 \\
Oracle Alice route probability & 714 & 0.2437 & 0.1723 & 0.0833 \\
\bottomrule
\end{tabular}
\end{table}

\begin{figure}[htb!]
\centering
\includegraphics[width=\linewidth]{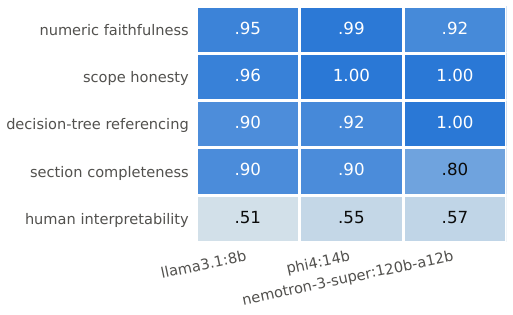}
\caption{Mean scores for local LLMs (columns). Rubric (rows) evaluation over $50$ scored prompt responses per LLM. Rubric criteria: numeric faithfulness (do stated numbers appear verbatim in the prompt), scope honesty (no security or physical-validity over-claims), decision tree referencing (does it refer to the correct tree leaf for its graph), and section completeness (how many of the requested sections are in the response), are automatically verified. The subjective, human interpretability score was evaluated by the authors and cross-referenced against Claude Fable~5 \cite{AnthropicFable52026} in an independent blind pass.}
\label{fig:llm_rubric}
\end{figure}

In Fig.~\ref{fig:llm_rubric}, we present scores for the LLM responses with 4 mechanical categories and 1 subjective human category. All LLMs scored near-perfect on the four mechanical checks, i.e. correctly reported the active decision tree leaf and its predicted value. Yet they were unable to inform \emph{why} the underlying features produced that outcome. No tested model produced responses exceeding a score of 7 for interpretability.

\section{Conclusion}

\subsection{Summary of results}
On a 50-topology corpus of SeQUeNCe E91 transcripts, Exp3 co-learning from pure bandit feedback recovers the minimax strategy structure and equilibrium support: bottlenecks are fatal, retention follows $1-1/N$, learned retention tracks the oracle reference at $r=0.99$, and both players' learned strategies match the coverage logic of the topology. The Nash gap of the averaged strategies decays with horizon, and the residual gap grows with action-space size---finite-horizon learning error, not topological weakness. Decision-tree faithfulness shows graph-level trees are nearly exact, Eve-action trees moderate, and Alice-route trees weak.

\subsection{Future work and open questions}

Open directions follow from the topology-corpus design: Alice's payoff is CHSH acceptance, Eve's payoff is route collision, and each matrix entry is estimated from cached SeQUeNCe trials. Improvements should distinguish better between denial, acceptance, and information leakage with more-explicit finite-sample and finite-key semantics.
The hardware fields are controlled simulation parameters, not emulated device physics. By using hardware studies only to motivate the parameter values, we create a non-trivial high-efficiency E91 routing regime. A future improvement should emulate device mappings more precisely. 
Finally, larger action spaces are less equilibrated over long training runs, and the residual Nash gap exposes horizon imbalance in the learned mixture. A natural next step is to test contextual-bandit variants \cite{Beygelzimer2011ContextualBandits} which exploit route and attack features, rather than treating all actions as uniformly exchangeable under exploration.

\section*{Acknowledgments and Code Availability}

We acknowledge funding by the Austrian Federal Ministry of Education, Science, and Research via the Austrian Research Promotion Agency (Forschungsf\"orderungsgesellschaft -- FFG) through Quantum Austria project No.\ 914033 and No.\ 63956271. This research was co-funded by the European Union (Quantum Flagship project ASPECTS, Grant Agreement No.\ 101080167). The authors acknowledge OpenAI Codex \cite{OpenAICodex2026} and Claude Fable~5 \cite{AnthropicFable52026} for repository development and manuscript editing assistance. 

The minimal reproducibility artifact for the topology corpus, payoff-cache construction, learning runs, and figure generation is available at the public GitHub repository \cite{Bell2026ExplainingQuantumNetworksCode}. The submitted experiments correspond to commit \texttt{ba96bb0}.



\bibliographystyle{IEEEtran}
\bibliography{references}

@IEEEtranBSTCTL{IEEEexample:BSTcontrol,
  CTLuse_forced_etal = {yes},
  CTLmax_names_forced_etal = {6},
  CTLnames_show_etal = {3}
}

@article{Ekert1991,
  title = {Quantum cryptography based on {Bell}'s theorem},
  author = {Ekert, Artur K.},
  journal = {Phys. Rev. Lett.},
  volume = {67},
  issue = {6},
  pages = {661--663},
  numpages = {0},
  year = {1991},
  month = {Aug},
  publisher = {American Physical Society},
  doi = {10.1103/PhysRevLett.67.661},
  url = {https://link.aps.org/doi/10.1103/PhysRevLett.67.661}
}

@article{Bell66,
  title = {On the Problem of Hidden Variables in Quantum Mechanics},
  author = {Bell, John S.},
  journal = {Rev. Mod. Phys.},
  volume = {38},
  issue = {3},
  pages = {447--452},
  numpages = {0},
  year = {1966},
  month = {Jul},
  publisher = {American Physical Society},
  doi = {10.1103/RevModPhys.38.447},
  url = {https://link.aps.org/doi/10.1103/RevModPhys.38.447}
}

@article{CHSH1969,
  author = {Clauser, John F. and Horne, Michael A. and Shimony, Abner and Holt, Richard A.},
  title = {Proposed experiment to test local hidden-variable theories},
  journal = {Physical Review Letters},
  volume = {23},
  number = {15},
  pages = {880--884},
  year = {1969},
  doi = {10.1103/PhysRevLett.23.880}
}

@article{Wu2021Sequence,
  author = {Wu, Xiaoliang and Kolar, Alexander and Chung, Joaquin and Jin, Dong and Zhong, Tian and Kettimuthu, Rajkumar and Suchara, Martin},
  title = {{SeQUeNCe}: A customizable discrete-event simulator of quantum networks},
  journal = {Quantum Science and Technology},
  volume = {6},
  number = {4},
  pages = {045027},
  year = {2021},
  doi = {10.1088/2058-9565/ac22f6}
}

@article{Coopmans2021NetSquid,
  author = {Coopmans, Tim and Knegjens, Robert and Dahlberg, Axel and Maier, David and Nijsten, Loek and de Oliveira Filho, Julio and Papendrecht, Martijn and Rabbie, Julian and Rozp{\k{e}}dek, Filip and Skrzypczyk, Matthew and Wubben, Leon and de Jong, Walter and Podareanu, Damian and Torres-Knoop, Ariana and Elkouss, David and Wehner, Stephanie},
  title = {{NetSquid}, a {NETwork} Simulator for {QUantum} Information using Discrete events},
  journal = {Communications Physics},
  volume = {4},
  pages = {164},
  year = {2021},
  doi = {10.1038/s42005-021-00647-8}
}

@article{BarrettKok2005,
  author = {Barrett, Sean D. and Kok, Pieter},
  title = {Efficient high-fidelity quantum computation using matter qubits and linear optics},
  journal = {Physical Review A},
  volume = {71},
  number = {6},
  pages = {060310},
  year = {2005},
  doi = {10.1103/PhysRevA.71.060310}
}

@article{Qi2007TimeShift,
  author = {Qi, Bing and Fung, Chi-Hang Fred and Lo, Hoi-Kwong and Ma, Xiongfeng},
  title = {Time-shift attack in practical quantum cryptosystems},
  journal = {Quantum Information and Computation},
  volume = {7},
  number = {1--2},
  pages = {73--82},
  year = {2007},
  doi = {10.26421/QIC7.1-2-3}
}

@article{Gisin2006Trojan,
  author = {Gisin, Nicolas and Fasel, Sylvain and Kraus, Barbara and Zbinden, Hugo and Ribordy, Gr{\'e}goire},
  title = {Trojan-horse attacks on quantum-key-distribution systems},
  journal = {Physical Review A},
  volume = {73},
  number = {2},
  pages = {022320},
  year = {2006},
  doi = {10.1103/PhysRevA.73.022320}
}

@article{Cui2026,
  author = {Cui, Z.-B. and Wang, Z.-Q. and Lai, P.-C. and Wang, Y. and Shi, J.-X. and Liu, P.-Y. and Sun, Y.-D. and Tian, Z.-C. and Liang, Y.-B. and Qi, B.-X. and Huang, Y.-Y. and Zhou, Z.-C. and Wu, Y.-K. and Xu, Y. and Duan, L.-M. and Pu, Y.-F.},
  title = {Metropolitan-scale ion-photon entanglement via a quantum network node with hybrid multiplexing enhancements},
  journal = {Nature Communications},
  volume = {17},
  pages = {697},
  year = {2026},
  doi = {10.1038/s41467-025-67311-5},
  note = {Published online 6 December 2025}
}

@article{Auer2002Exp3,
  author = {Auer, Peter and Cesa-Bianchi, Nicol{\`o} and Freund, Yoav and Schapire, Robert E.},
  title = {The nonstochastic multiarmed bandit problem},
  journal = {SIAM Journal on Computing},
  volume = {32},
  number = {1},
  pages = {48--77},
  year = {2002},
  doi = {10.1137/S0097539701398375}
}

@article{FreundSchapire1999,
  author = {Freund, Yoav and Schapire, Robert E.},
  title = {Adaptive game playing using multiplicative weights},
  journal = {Games and Economic Behavior},
  volume = {29},
  number = {1--2},
  pages = {79--103},
  year = {1999},
  doi = {10.1006/game.1999.0738}
}

@book{Breiman1984CART,
  author = {Breiman, Leo and Friedman, Jerome H. and Olshen, Richard A. and Stone, Charles J.},
  title = {Classification and Regression Trees},
  publisher = {Wadsworth International Group},
  address = {Belmont, CA},
  year = {1984},
  isbn = {0-534-98053-8}
}

@article{Rudin2019Interpretable,
  author = {Rudin, Cynthia},
  title = {Stop explaining black box machine learning models for high stakes decisions and use interpretable models instead},
  journal = {Nature Machine Intelligence},
  volume = {1},
  number = {5},
  pages = {206--215},
  year = {2019},
  doi = {10.1038/s42256-019-0048-x}
}

@inproceedings{BaileyPiliouras2018MWU,
  author = {Bailey, James P. and Piliouras, Georgios},
  title = {Multiplicative Weights Update in Zero-Sum Games},
  booktitle = {Proceedings of the 2018 ACM Conference on Economics and Computation},
  series = {EC '18},
  pages = {321--338},
  year = {2018},
  publisher = {Association for Computing Machinery},
  address = {New York, NY, USA},
  doi = {10.1145/3219166.3219235}
}

@article{WashburnWood1995,
  author = {Washburn, Alan and Wood, Kevin},
  title = {Two-Person Zero-Sum Games for Network Interdiction},
  journal = {Operations Research},
  volume = {43},
  number = {2},
  pages = {243--251},
  year = {1995},
  doi = {10.1287/opre.43.2.243}
}

@article{Satoh2021,
  author = {Satoh, Takahiko and Nagayama, Shota and Suzuki, Shigeya and Matsuo, Takaaki and Hajdu{\v{s}}ek, Michal and Van Meter, Rodney},
  title = {Attacking the Quantum Internet},
  journal = {IEEE Transactions on Quantum Engineering},
  volume = {2},
  pages = {1--17},
  year = {2021},
  doi = {10.1109/TQE.2021.3094983},
  note = {Art. no. 4102617}
}

@inproceedings{Beygelzimer2011ContextualBandits,
  author = {Beygelzimer, Alina and Langford, John and Li, Lihong and Reyzin, Lev and Schapire, Robert E.},
  title = {Contextual Bandit Algorithms with Supervised Learning Guarantees},
  booktitle = {Proceedings of the Fourteenth International Conference on Artificial Intelligence and Statistics},
  series = {Proceedings of Machine Learning Research},
  volume = {15},
  pages = {19--26},
  year = {2011}
}

@misc{OllamaSoftware2026,
  author       = {{Ollama}},
  title        = {{Ollama}},
  year         = {2026},
  howpublished = {Software},
  url          = {https://github.com/ollama/ollama},
  note         = {Accessed Jul. 10, 2026}
}

@misc{AnthropicFable52026,
  author = {{Anthropic}},
  title = {Claude Fable 5 and Claude Mythos 5},
  howpublished = {Large language model},
  year = {2026},
  url = {https://www.anthropic.com/news/claude-fable-5-mythos-5},
  note         = {Accessed 2026-07-10}
}

@misc{OllamaNemotron3Super,
  author       = {{Ollama}},
  title        = {{nemotron-3-super:120b}},
  year         = {2026},
  howpublished = {Ollama model library},
  url          = {https://ollama.com/library/nemotron-3-super:120b},
  note         = {Accessed 2026-07-10}
}

@misc{OllamaPhi4,
  author       = {{Ollama}},
  title        = {{phi4:14b}},
  year         = {2024},
  howpublished = {Ollama model library},
  url          = {https://ollama.com/library/phi4:14b},
  note         = {Accessed 2026-07-10}
}

@misc{OllamaLlama31,
  author       = {{Ollama}},
  title        = {{llama3.1:8b}},
  year         = {2024},
  howpublished = {Ollama model library},
  url          = {https://ollama.com/library/llama3.1:8b},
  note         = {Accessed 2026-07-10}
}

@misc{Bell2026ExplainingQuantumNetworksCode,
  author = {Bell, Brennan},
  title = {Explaining Quantum Networks},
  year = {2026},
  howpublished = {GitHub repository},
  url = {https://github.com/496crows/explaining-quantum-networks},
  note         = {Accessed 2026-07-10}
}

@misc{OpenAICodex2026,
  author       = {{OpenAI}},
  title        = {{Codex}: AI Coding Agents for Software Engineering},
  year         = {2026},
  howpublished = {\url{https://openai.com/codex/}},
  note         = {Accessed 2026-07-10}
}
\end{document}